\input harvmac

%%%%%%%%%%%%%%%%%%%%%%%%%%%%%%%%%%%%%%%%%%%%%%%%%%%%%%%%%%%%%%%%%%%%%

% MACRO for REF.

\def\np#1#2#3{Nucl. Phys. {\bf B#1}, (19#2) #3}
\def\ibid#1#2#3{{\it ibid.} {\bf B#1}, (19#2) #3}
\def\pln#1#2#3{Phys. Lett. {\bf B#1}, (19#2) #3}
\def\plo#1#2#3{Phys. Lett. {\bf #1B}, (19#2) #3}
\def\pr#1#2#3{Phys. Rev. {\bf D#1}, (19#2) #3}

\def\hp#1{{\tt hep-th/#1}}
\def\hpp#1{{\tt hep-ph/#1}}

% MACRO for TEXT

\def\vev#1{\langle #1 \rangle}
\def\EU#1{\epsilon^{#1}}
\def\ED#1{\epsilon_{#1}}
\def\QE#1{Q_{{\rm e}[#1]}}
\def\QM#1{Q_{{\rm m}[#1]}}
\def\Pf{{\rm Pf}}

%%%%%%%%%%%%%%%%%%%%%%%%%%%%%%%%%%%%%%%%%%%%%%%%%%%%%%%%%%%%%%%%%%%%%
% LIST of Reference

\lref\DSB{
E.~Witten, 
\np{188}{81}{513}.
}
\lref\index{
E.~Witten, 
\np{202}{82}{253}.
}
\lref\seiberg{
N.~Seiberg,
RU-94-64, \hp{9408013};
RU-95-37, \hp{9506077}.
}
\lref\lecture{
K.~Intriligator and N.~Seiberg,
RU-95-48, IASSNS-HEP-95/70, \hp{9509066}.
}
\lref\ADS{
I.~Affleck, M.~Dine and N.~Seiberg,
\np{241}{84}{493};
\ibid{256}{85}{557}.
}
\lref\AKMRV{
D.~Amati, K.~Konishi, Y.~Meurice, G.C.~Rossi and G.~Veneziano,
{\it Phys. Rept.} {\bf 162} (1988) 169.
}
\lref\simple{
K.~Intriligator, N.~Seiberg and S.H.~Shenker,
\pln{342}{95}{152}, \hpp{9410203}.
}
\lref\spin{
P.~Pouliot,
\pln{359}{95}{108}, \hp{9507018}.
}
\lref\SOten{
I.~Affleck, M.~Dine and N.~Seiberg,
\plo{140}{84}{59}.
}
\lref\variety{
M.A.~Luty and W.~Taylor IV,
MIT-CTP-2440, \hp{9506098}.
}
\lref\tHooft{
G.~'t Hooft,
in {\it Recent Developments in Gauge Theories,}
eds. G.~'t Hooft {\it et al.}
( Plenum Press, New York, 1980 )
}

%%%%%%%%%%%%%%%%%%%%%%%%%%%%%%%%%%%%%%%%%%%%%%%%%%%%%%%%%%%%%%%%%%%%%

\Title{\vbox{\hbox{    YITP-96-5   }
\hbox{\tt           hep-th/9602035 } 
}}
{\vbox{
\centerline{
       Duality of $N=1$ Supersymmetric $SO(10)$ Gauge Theory
}
\centerline{
          with Matter in the Spinorial Representation 
}
}}

\vskip .2in

\centerline{
                        Teruhiko Kawano 
}

\vskip .2in 

\centerline{
            Yukawa Institute for Theoretical Physics, 
}
\centerline{
             Kyoto University, Kyoto 606-01, Japan
}
\vskip .1in
\centerline{\tt{
                  kawano@yukawa.kyoto-u.ac.jp
}}

\vskip .3in
We study $N=1$ supersymmetric $SO(10)$ gauge theory 
with a field in the spinorial representaition and $N_f$ 
($\leq8$) fields in the defining representation. 
It is shown that this theory
for $N_f=7,8$ has a dual description, which is $N=1$ supersymmetric 
$SU(N_f-5)$ gauge theory. 
Its matter content for $N_f=7$ is different {}from the one for $N_f=8$; 
for $N_f=7$, it contains $8$ fields in the anti-fundamental representation. 
For $N_f=8$, a rank-$2$ symmetric tensor and one field in the fundamental 
representation appears in addition to them.
This duality connects along the flat direction 
to the duality between chiral and vector gauge theory 
found by Pouliot.

\Date{February, 1996}

%\newsec{Introduction}

The dynamical supersymmetry breaking (DSB) is a candidate to resolve
the gauge hierarchy problem\DSB. In the minimal SUSY standard model, 
the electroweak symmetry cannot be broken without the soft SUSY breaking
terms. Thus the DSB may explain why the electroweak scale is so small
compared with the Planck scale. However the Witten index\index~tells
us that it is not easy to find models where the DSB can occur.

It has been discussed\SOten~that the DSB occurs in $SO(10)$ gauge theory 
with a matter in the spinorial representation {\bf 16}. 
Since this model is strongly interacting, the argument is indirect\SOten. 
This is why this model is called non-calculable model. 
(for other example, see 
\nref\SUfive{
I.~Affleck, M.~Dine and N.~Seiberg,
\plo{137}{84}{187}.
}
\nref\murayama{
H.~Murayama,
\pln{355}{95}{187}, \hp{9505082}.
}
\nref\PopTri{
E.~Poppitz and S.P.~Trivedi, 
FERMILAB-PUB-95-258-T-REV, \hp{9507169}.
}
\nref\pouliot{
P.~Pouliot,
RU-95-66, \hp{9510148}.
}
\hskip -5mm \refs{\SUfive-\pouliot}.)
Recently Murayama\murayama~has introduced a field in the
defining representation {\bf 10} into this model to have flat
directions and has shown that the DSB occurs in this model.
The presence of flat directions makes it easier to study 
the low energy behavior of $N=1$ supersymmetric gauge theories, 
since we can apply the technique developed by Seiberg 
and his collaborators
\nref\versus{
N.~Seiberg,
\pln{318}{93}{469}, \hpp{9309335}.
}
\nref\exactW{
K.~Intriligator, R.G.~Leigh and N.~Seiberg,
\pr{50}{94}{1092}, \hp{9403198}.
}
\nref\moduli{
N.~Seiberg,
\pr{49}{94}{6857}, \hp{9402044}.
}
\nref\duality{
N.~Seiberg,
\np{435}{95}{129}, \hp{9411149}.
}
\hskip -5mm \refs{\versus-\duality}.
These technique has been revealing the non-perturbative
properties of $N=1$ supersymmetric gauge theories. 
(see \seiberg, \lecture~and references therein.)
Thus, after giving mass to the field in the defining representation and 
decoupling it, we can see that the DSB occurs in the original model, 
if we assume no phase transition occurs.

In this paper, we study the quantum moduli space of this model 
into which $N_f$ fields in the defining representation are introduced.
It is shown that the low energy theory of this model for $N_f=7,8$ 
has a dual description, which is $SU(N_f-5)$ gauge theory.
Its matter content for $N_f=7$ is different {}from the one for $N_f=8$; 
for $N_f=7$, it contains $8$ fields in the anti-fundamental representation.
For $N_f=8$, a rank-$2$ symmetric tensor and one field in the fundamental 
representation appears in addition to them.
This duality connects along the flat direction 
to the duality between $SU(N_f-5)$ chiral and $SO(7)$ vector 
gauge theory found by Pouliot\spin.

Hopefully this result may shed some light toward
the solution for the gauge hierarchy problem.

%\newsec{SO(10) gauge theory with a spinorial rep. 
%and $N_f$ defining reps.}
\break
\hskip -5mm {\bf SO(10) gauge theory with a spinorial rep. 
and $N_f$ defining reps.}

We consider $N=1$ supersymmetric $SO(10)$ gauge theory 
with a field $\psi_{\alpha}$ $(\alpha=1,\cdots,16)$ 
in the spinorial representation {\bf 16} 
and $N_f$ fields $H^i_A$ $(A=1,\cdots,10; i=1,\cdots,N_f)$ 
in the defining representation {\bf 10}.
This theory is asymptotically free, if $N_f < 22$.
The global symmetries are $SU(N_f)\times U(1)_F\times U(1)_R$.
In our convention,
the fields transform under these $U(1)$ symmetries in the following way:
\eqn\Rsymm{\eqalign{
U(1)_R :
&~~\psi_\alpha(\theta) \rightarrow \psi_\alpha(e^{-i\omega}\theta),
\cr         
&~~H^i_A(\theta) \rightarrow e^{i{N_f-6 \over N_f}\omega}
                                         H^i_A(e^{-i\omega}\theta).
\cr
U(1)_F :
&~~\psi_\alpha(\theta) \rightarrow e^{-iN_f\omega}\psi_\alpha(\theta),
\cr         
&~~H^i_A(\theta) \rightarrow e^{2i\omega}H^i_A(\theta).
\cr}}

\bigskip
\hskip -5mm {\it 1. $N_f=8$ and Magnetic $SU(3)$ gauge theory}

We begin to consider the low energy effective theory for $N_f=8$. 
All the gauge invariant operators can be generated by the operators
\eqn\Gop{\eqalign{
M^{ij} &= H^i_AH^j_A, 
\cr
Y^i &= \psi^c\gamma_A\psi H^i_A,
\cr
\QE{i_1i_2i_3} &= \epsilon_{i_1i_2i_3j_1\cdots j_5} 
H^{j_1}_{A_1}\cdots H^{j_5}_{A_5}
\psi^c\gamma_{A_1\cdots A_5}\psi,
\cr}}
where $\psi^c$ is the charge conjugation of $\psi$.
In absense of any superpotential at tree level, 
the flat directions are parametrized by these operators \ADS,
\variety~with classical constraints  

\eqn\constviii{\eqalign{
\QE{i_1i_2i_3}\QE{j_1j_2j_3}
= & -{1\over 6}Y^mM^{k_1l_1}M^{k_2l_2}   
\bigg[  \ED{mk_1k_2i_1i_2i_3j_1j_2}\QE{j_3l_1l_2}
\cr
&~~~~~~+ (\hbox{\rm permutation of}~j_1, j_2~\hbox{\rm and}~j_3 ) \bigg]
\cr
& +{1\over 5!}\ED{i_1i_2i_3k_1\cdots k_5}
\ED{j_1j_2j_3l_1\cdots l_5}M^{k_1l_1}\cdots M^{k_4l_4}Y^{k_5}Y^{l_5},
\cr
Y^i\QE{ij_1j_2} = & 0.
\cr}}

These constraints are not modified by any quantum corrections. 
This low energy spectrum $M^{ij}$, $Y^i$ and $Q_{\rm e}$ does not 
saturate the 't Hooft anomaly matching conditions\tHooft.
In the quantum theory, the low energy effective superpotential
cannot be written in terms of $M^{ij}$, $Y^i$ and $Q_{\rm e}$ alone 
without any singularities. We suspect {}from these facts 
that there appear new light degrees
of freedom, as SUSY QCD for $N_f\geq N_c+2$~\duality.

In fact, though it is a conjecture, this quantum theory has 
a dual description which we call `magnetic theory'; 
$SU(3)$ gauge theory with a symmetric tensor $S_{ab}$ in {\bf 6}, 
a field $q_a$ in the fundamental representation {\bf 3}, 
$8$ fields $\bar q^a_i$ in the anti-fundamental representation 
$\bar {\bf 3}$ and singlets $M^{ij}$, $Y^i$.
The superpotential in the magnetic theory is 
\eqn\magW{
W_{\rm m} =  {1\over \mu_1^2} M^{ij}\bar q^a_i\bar q^b_jS_{ab}
      +{1\over \mu_2^2} Y^i\bar q^a_i q_a
      +\det S,
}
where dimensionful parameters $\mu_1$ and $\mu_2$ are 
introduced to give the `meson' fields 
$M^{ij}$, $Y^i$ the same mass dimensions as those in the `electric' theory.
This theory is asymptotically free.
Under the global symmetries $SU(N_f)\times U(1)_F\times U(1)_R$, 
the fields $S_{ab}$, $q_a$, $\bar q^a_i$, $M^{ij}$ and $Y^i$ 
transform as $({\bf 1},0,{2\over3})$, $({\bf 1}, 16, {4\over3})$, 
$(\bar {\bf 8}, -2, {5\over12})$, $({\bf 36}, 4, {1\over2})$ 
and $({\bf 8}, -14, {1\over4})$, respectively.

As a consistency check on this duality, the 't Hooft anomaly
matching conditions\tHooft~are satisfied.
The gauge invariant operators $M^{ij}$, $Y^i$ and $\QE{i_1i_2i_3}$ in the 
electric $SO(10)$ theory correspond to the operators $M^{ij}$, $Y^i$ 
and $\QM{i_1i_2i_3}=\ED{a_1a_2a_3}\bar q^{a_1}_{i_1}
\bar q^{a_2}_{i_2} \bar q^{a_3}_{i_3}$ in the magnetic $SU(3)$ theory.
The other gauge invariant operators in the magnetic theory can be 
written by combining the above operators $M^{ij}$, $Y^i$ and $Q_{\rm m}$
or identically vanish, through the equations of motion 
{}from the superpotential $W_{\rm m}$.

At the point $\vev{M^{88}}\not=0$ and $\vev{Y^{8}}\not=0$ on the
moduli space, the electric $SO(10)$ gauge group is broken to $SO(7)$ 
by the vacuum expectation value of $H^8_A$ and $\psi_{\alpha}$, and 
the other $7$ fields $H^i_A$ in the defining representation turn into 
those in the spinorial representation ${\bf 8}$ under the unbroken $SO(7)$. 
(see the appendix in \murayama.)
On the other hand, in the magnetic theory, 
the gauge group $SU(3)$ remains unbroken and  
the fields $\bar q^a_8$ and $q_a$ decouple at the energy scale below 
${1\over\mu_2^2}\vev{Y^8}$. Solving the equations of motion, 
we obtain the low energy superpotential 
$W_{\rm eff}=\sum^7_{i,j=1}{1\over\mu_1^2}M^{ij}
\bar q_i^a\bar q_j^bS_{ab} + \det S$.
Thus the present system just reduces to Pouliot's model\spin~in which 
the duality $SO(7)\leftrightarrow SU(3)$ was found.
This is a non-trivial check on the duality of the model 
under consideration. 

\bigskip
\hskip -5mm {\it 2. $N_f=7$}

For $N_f=7$, the gauge invariant operators are the same as those for 
$N_f=8$ except for $Q_{\rm e}$; 
$\QE{i_1i_2}= \epsilon_{i_1i_2j_1\cdots j_5} 
H^{j_1}_{A_1}\cdots H^{j_5}_{A_5}
\psi^c\gamma_{A_1\cdots A_5}\psi,$
instead of $\QE{i_1i_2i_3}$. The classical constraints 
among these operators are
\eqn\constvii{\eqalign{
\QE{i_1i_2}\QE{j_1j_2}=
&-{1\over6}Y^m\ED{mi_1i_2j_1j_2k_1k_2}\QE{l_1l_2}M^{k_1l_1}M^{k_2l_2}
\cr
&+{1\over5!}\ED{i_1i_2k_1\cdots k_5}\ED{j_1j_2l_1\cdots \l_5}
M^{k_1l_1}\cdots M^{k_4l_4}Y^{k_5}Y^{l_5},
\cr
Y^i\QE{ij}=&0.
\cr}}

These constraints remain to hold even in the quantum theory. 
The light degrees of freedom $M^{ij}$, $Y^i$ and $Q_{\rm e}$ again 
do not saturate the 't Hooft anomaly matching conditions\tHooft. 

In the electric theory for $N_f=8$, after giving a mass to the field 
$H^8_A$ and decoupling it, we have the electric theory for $N_f=7$. 
Alternatively, assuming the above duality for $N_f=8$, 
adding a mass term $mM^{88}$ to the magnetic superpotential 
\magW~and using the equations of motion, 
we can see that the gauge symmetry $SU(3)$ breaks to $SU(2)$ 
by $\vev{\bar q^3_8}\not=0$ and 
the symmetric tensor $S_{ab}$ become massive.
After integrating out the field $S_{ab}$, we find the result 
proportional to $M^{ij}M^{kl}\bar q^a_i\bar q^b_k\ED{ab}
\bar q^c_j\bar q^d_l\ED{cd}$ in the superpotential.
In addition, by the holomorphy and the symmetries\versus, the term 
$\det M({1\over M})_{kl}Y^kY^l$ can be induced in the superpotential.
Summing up these contributions, we
obtain the following low energy superpotential:
\eqn\Wvii{\eqalign{
W_{\rm eff}=&{1\over\Lambda_7^{15}}
\bigg[\det M\left({1\over M}\right)_{kl}Y^{k}Y^{l}
-{\mu^{10}\over2!}M^{ij}M^{kl}\bar q^a_i\bar q^b_k\ED{ab}
\bar q^c_j\bar q^d_l\ED{cd}\bigg]
\cr
&+{1\over\mu_2^2}Y^i\bar q^a_iq_a
\cr}}
with $i, j, k$ and $l$ now running {}from $1$ to $7$
and $a, b, c$ and $d$ running {}from $1$ to $2$.
The scale $\Lambda_7$ is the dynamical scale in the electric
theory, which is connected to the dynamical scale 
$\tilde\Lambda_{2,7}$ in the magnetic theory as 
$\Lambda_7^{15}[\tilde\Lambda^2_{2,7}]^2=
\vev{S_{33}}^3{\mu_1}^{12}{\mu_2}^4$. The dimensionful parameter 
$\mu$ is defined by 
${\mu}^5=-\vev{S_{33}}\mu_1^4\mu_2^2/\tilde\Lambda^2_{2,7}$.
The composite operator $\QM{ij}=\mu^5\bar q^a_i\bar q^b_j\ED{ab}$ 
can be identified with $\QE{ij}$ in the electric theory. 
The operator $\bar q^a_iq_a$ is vanishing as we can see 
{}from the equation of motion for $Y^i$ and the other gauge invariant 
operators can be constructed by combining the operators 
$M^{ij}$, $Y^i$ and $Q_{\rm m}$. Thus the correspondance of the gauge
invariant operators between the electric and the magnetic theory 
remains for $N_f=7$. {}From the equations of motion for $M^{ij}$ 
and $q_a$, the constraints\constvii~in the electric theory are 
partly reproduced {}from the magnetic superpotential\Wvii. 

This theory is asymptotically free.
(Note that it is not asymptotically free until integrating out the
symmetric tensor $S_{ab}$.)
We can verify that these fields saturate the 't Hooft anomaly matching
conditions\tHooft. 

At the point $\vev{M^{77}}\not=0$, $\vev{Y^7}\not=0$ on the moduli
space, the electric gauge group $SO(10)$ breaks to $SO(7)$ and the
matter content becomes $6$ fields in the spinorial representation.
For the magnetic theory, on the other hand, 
$\bar q^a_7$ and $q_a$ become massive and should
be integrated out. By the holomorphy and the symmetries\versus, there is 
another allowed contribution proportional to 
$\EU{7k_1\cdots k_6}\bar q^{a_1}_{k_1}\bar q^{a_2}_{k_2}\ED{a_1a_2}
\cdots \bar q^{a_5}_{k_5}\bar q^{a_6}_{k_6}\ED{a_5a_6}$.
The resultant superpotential is proportional to 
$\det M - {1\over2}M^{ij}M^{kl}B_{ik}B_{jl} - \Pf~B$
with $B_{ij}\equiv\QM{ij}/\vev{Y^7}$.
This is exactly the same superpotential as was found by Pouliot\spin, 
who has shown that these composite fields $M^{ij}$ and $B_{ij}$ 
saturate the 't Hooft anomaly matching conditions\tHooft~for the microscopic 
$SO(7)$ gauge theory. Conversely the equations of motion 
{}from this superpotential reproduce the classical 
constraints \constvii~in the electric theory.
This is interesting because the classical relations 
in the electric theory is the result by
the non-perturbative dynamics in the magnetic theory.
Therefore the electric $SO(10)$ gauge theory 
and the magnetic $SU(2)$ reduce to the same low energy theory in the
infrared at this point $\vev{M^{77}}\not=0$, $\vev{Y^7}\not=0$ 
on the moduli space.

Thus our conjecture follows that the $SU(2)$ gauge theory with 
the fields $\bar q^a_i$, $q_a$, $M^{ij}$ and $Y^i$ and the above
superpotential \Wvii~describes the low energy dynamics 
of the $SO(10)$ gauge theory for $N_f=7$.

\bigskip
\hskip -5mm {\it 3. $1\leq N_f\leq 6$}

For $N_f=6$, the flat directions can be described by the gauge
invariant operators $M^{ij}$, $Y^i$ and 
$Q_{i}=\ED{ij_1\cdots j_5}\psi^c\gamma_{A_1\cdots A_5}\psi 
H^{j_1}_{A_1}\cdots H^{j_5}_{A_5}$, as we have seen for $N_f=7$ and $8$.
The classical constraints are
\eqn\constvi{\eqalign{
Q_{i}Q_{j} &= {1\over5!}\ED{ik_1\cdots k_5}\ED{jl_1\cdots l_5}
M^{k_1l_1}\cdots M^{k_4l_4}Y^{k_5}Y^{l_5},
\cr
Y^iQ_{i}&=0.
\cr}}

By $U(1)_R$ symmetry, no effective superpotential is dynamically 
generated. The quantum theory has the moduli space of the
degenerate ground states. The massless spectrum $M^{ij}$, $Y^i$ and 
$Q_i$ does not saturate the 't Hooft anomaly matching conditions\tHooft.
Symmetry argument suggests
that the classical constraints \constvi~have the
possibility to be modified quantum mechanically into 
\eqn\Qconstvi{\eqalign{
\det M \left({1\over M}\right)_{kl}Y^kY^l- M^{kl}Q_{k}Q_{l}
&=\Lambda_6^{16},  
\cr
Y^iQ_{i}&=0,
\cr}}
with $\Lambda_6$ being the dynamical scale in the electric theory
for $N_f=6$.

We show consistency checks on this quantum constraints \Qconstvi.
First, along the flat directions $\vev{M^{66}}\not=0$, $\vev{Y^6}\not=0$
by which the electric $SO(10)$ gauge theory turns into the $SO(7)$ gauge
theory with $5$ spinorial representations, 
the quantum constraints \Qconstvi~reduce to 
$\det M - M^{ij}B_iB_j = \Lambda_{SO(7)}^{10}$
with $i$, $j$ running {}from $1$ to $5$,
which is the quantum constraint found for $N_f=5$ in \spin, 
if we identify $B_i=Q_i/\vev{Y^6}$ and 
$\Lambda_{SO(7)}^{10}=\Lambda_6^{16}/\vev{Y^6}^2$.
Second, adding a mass term $mM^{77}$ to the superpotential \Wvii~in 
the dual magnetic theory and integrating out massive fields, 
we find the magnetic gauge group $SU(2)$ is completely broken 
by $\vev{\bar q^2_7}\not=0$ and obtain the quantum 
constraints \Qconstvi~under the one-loop matching condition 
$\Lambda_6^{16}=m\Lambda_7^{15}$.
Here the operator $Q_i$ turns out to be
$\mu^5\vev{\bar q^2_7}\bar q^1_i$.

The quantum constraints \Qconstvi~are implemented by the following 
effective superpotential for $N_f=6$:
\eqn\Wvi{
W_{\rm eff}= X\left(\det M \left({1\over M}\right)_{kl}Y^kY^l-M^{kl}Q_kQ_l
-\Lambda_6^{16}\right) + LY^iQ_i
}
where $X$ and $L$ are the Lagrange multiplier fields.

For $N_f=5$, by adding a mass term $mM^{66}$ to the above
superpotential \Wvi~and integrating out massive fields, we can find
the low energy effective superpotential for $N_f=5$ 
\eqn\Wv{
W_{\rm eff} = \left[ {\Lambda_5^{17} \over (M^4\cdot Y^2)-Q^2} \right]
}
with the scale $\Lambda_5^{17}=m\Lambda_6^{16}$and the operator
$Q=Q_6$, where 
$(M^4\cdot Y^2)=\det M({1\over M})_{kl}Y^kY^l$.
The singularity at the point $(M^4\cdot Y^2)-Q^2=0$ indicates that 
the gauge symmetry is not completely broken and the subgroup
remains at this point.

By the $SO(7)$ flat directions $\vev{M^{55}}\not=0$,
$\vev{Y^5}\not=0$, the superpotential \Wv~reproduces
the effective superpotential $\Lambda_{SO(7)}^{11}/[\det M - B^2]$ 
for $N_f=4$ in \spin~ 
under the identification $B=Q/\vev{Y^5}$ and
$\Lambda_{SO(7)}^{11}=\Lambda_5^{17}/\vev{Y^5}^2$.

For $1\leq N_f \leq 4$, 
by the holomorphy and the symmetries\versus, the following low 
energy effective superpotential $W_{\rm eff}$ is allowed to arise:  
\eqn\Weff{
W_{\rm eff} = (6-N_f)\, \left[ {\Lambda^{22-N_f}_{N_f} 
\over (M^{N_f-1}\cdot Y^2)}\right]^{1 \over 6-N_f},
}
where $(M^{N_f-1}\cdot Y^2)=\det M({1\over M})_{kl}Y^kY^l$.
Indeed, adding mass terms and decoupling massive
fields, we find that these coefficients are consistently determined 
under the one-loop matching condition 
$\Lambda_{N_f}^{22-N_f}=m\Lambda_{N_f+1}^{21-N_f}$. 
Furthermore by adding a mass term $mM^{55}$ to the low energy 
superpotential \Wv~for $N_f=5$ and integrating out massive fields, 
we obtain the superpotential \Weff~for $N_f=4$.

Thus the theory for $1\leq N_f\leq 5$ has no vacuum, 
until we properly add superpotentials at tree level to it.

\bigskip
\centerline{{\bf Acknowledgements}}

I would like to thank K.~Inoue, Y.~Kikukawa, N.~Maekawa, M.~Ninomiya 
and S.~Sugimoto for comments, discussions and their encouragement. 
I am most grateful to T.~Kugo for very valuable comments 
and suggestions on the manuscript.
This work was supported in part by JSPS Research Fellowships for Young
Scientists.

{\bf Note added:} Upon completion of this manuscript, I received 
the paper\ref\spinten{
P~.Pouliot and M.J.~Strassler, 
RU-95-78, \hp{9602031}. 
}, where their results are consistent with ours.

\listrefs

\end